\documentclass[
 aps, pra,
 amsmath,amssymb,
 11pt,
 final,
tightenlines,
 twoside,
 twocolumn,
 nofloats,
nofootinbib,
 superscriptaddress,
showkeys,
showkeywords,
 ]
{revtex4-2}

\usepackage[T2A]{fontenc}
\usepackage[utf8]{inputenc}
\usepackage[english]{babel}
\usepackage{graphicx}
\usepackage{dcolumn}
\usepackage{bm}

\usepackage{xcolor}
\usepackage{soul}
\usepackage{changes}

\input{maik.rty}

\setcitestyle{authoryear,round}
\def\squareforqed{\hbox{\rlap{$\sqcap$}$\sqcup$}}

\def\sq{\ifmmode\squareforqed\else{\unskip\nobreak\hfil
\penalty50\hskip1em\null\nobreak\hfil\squareforqed
\parfillskip=0pt\finalhyphendemerits=0\endgraf}\fi}

\def\degr{\hbox{$^\circ$}}

\def\utw{\smash{\rlap{\lower5pt\hbox{$\sim$}}}}

\def\udtw{\smash{\rlap{\lower6pt\hbox{$\approx$}}}}

\def\fm{\hbox{$\,.\!\!^{\rm m}$}}

\def\fdg{\hbox{$\,.\!\!^\circ$}}

\def\diameter{{\ifmmode\mathchoice
{\ooalign{\hfil\hbox{$\displaystyle/$}\hfil\crcr
{\hbox{$\displaystyle\mathchar"20D$}}}}
{\ooalign{\hfil\hbox{$\textstyle/$}\hfil\crcr
{\hbox{$\textstyle\mathchar"20D$}}}}
{\ooalign{\hfil\hbox{$\scriptstyle/$}\hfil\crcr
{\hbox{$\scriptstyle\mathchar"20D$}}}}
{\ooalign{\hfil\hbox{$\scriptscriptstyle/$}\hfil\crcr
{\hbox{$\scriptscriptstyle\mathchar"20D$}}}}
\else{\ooalign{\hfil/\hfil\crcr\mathhexbox20D}}%
\fi}}

\begin{document}



\title{On the causes of brightness variability of the young star BP~Tau.}

\author{\firstname{M.~A.}~\surname{Burlak}}
\affiliation{Sternberg Astronomical Institute, M.V. Lomonosov Moscow State University, Moscow, 119234 Russia\\}
\author{\firstname{K.~N.}~\surname{Grankin}}
\affiliation{Crimean Astrophysical Observatory RAS, 298409 Nauchny, Republic of Crimea \\}
\author{\firstname{A.~V.}~\surname{Dodin}}
\author{\firstname{A.~V.}~\surname{Zharova}}
\author{\firstname{N.~P.}~\surname{Ikonnikova}}
\author{\firstname{V.~A.}~\surname{Kiryukhina}}
\author{\firstname{S.~A.}~\surname{Lamzin}}
\email{lamzin@sai.msu.ru}
\author{\firstname{B.~S.}~\surname{Safonov}}
\author{\firstname{I.~A.}~\surname{Strakhov}}
\affiliation{Sternberg Astronomical Institute, M.V. Lomonosov Moscow State University, Moscow, 119234 Russia\\}

Accepted by Astrophysical Bulletin


\begin{abstract}
We have constructed and analysed the secular light curve of BP~Tau, a classical T~Tauri-type star. Wave-like variations in the average brightness were detected, with an amplitude of $\Delta B\approx 0\fm 2$ and characteristic time-scales of several decades. We argue that three deep dimming events $(\Delta B \sim 1\fm 5)$, lasting from 1 hour to several days, are caused by the eclipse of a hot (accretion) spot by dust falling onto the star together with gas. Such eclipses, albeit with smaller amplitudes, may explain the absence of a strictly defined periodicity in the brightness variations of BP Tau associated with axial rotation. We also show that within the distance range of 0.1 to 200 AU, BP Tau does not have a companion with a mass exceeding 0.2~M$_\odot.$ The causes of brightness and colour index variations on different time-scales are discussed.
\end{abstract}
\maketitle

\section{Introduction}
 \label{sect:introduct}
 
T~Tauri stars (TTS) are young stars (with ages $t \lesssim 10^7$\,yr) having masses $\lesssim 2$\,M$_\odot$ that are in the process of contracting towards the main sequence. T~Tauri stars are conventionally divided into two subgroups \citep{Bertout-1989}: classical T~Tauri stars (CTTS) and weak-line T~Tauri stars (WTTS). In the case of WTTS, emission in certain lines within the visible and ultraviolet ranges, the presence of cool spots $(T_{\text sp} < T_\text{eff})$, and intense X-ray emission are attributed to chromospheric and coronal activity similar to solar-type activity but much more powerful. In contrast, the activity of CTTS — including variability across all spectral ranges, line and continuum emission, outflows from the stellar vicinity, and more — is primarily associated with accretion from the protoplanetary disc \citep{Hartmann-2016}. Both accretion-driven and chromospheric-coronal activity in TTS are fundamentally governed by the strong magnetic fields of these stars. For instance, near the surface of both CTTS and WTTS, the dipole component of their global magnetic field has a strength of $\gtrsim 1$\,kG \citep{Donati-2009}.

In addition to cool spots, CTTS also exhibit hot spots $(T_{\text sp} > T_\text{eff})$, which form in regions where the accretion stream collides with the stellar surface \citep{1980Afz....16..243G}. Hot spots cover approximately 10\,\%{} of the surface of CTTS \citep{Hartmann-2016}. Due to rotational modulation, this should lead to periodic variability in the brightness of these objects, at least when their rotation axis is inclined sufficiently relative to the line of sight \citep{Herbst-94}.

Such periodicity has been observed in many CTTSs, however, it is absent in some stars \citep[see][and references therein]{Rebull-2020, Cody-2022}, 
probably because the parameters and/or positions of the hot spots on the stellar surface vary over time-scales comparable to the rotation period. Three-dimensional MHD modelling of the magnetospheric accretion process shows that such phenomena arise as a consequence of the non-stationary nature of the interaction between the inner regions of the disc and the magnetic field of the young star under specific combinations of their parameters. The characteristic time-scale for the manifestation of this non-stationarity ranges from several hours to several days \citep{Romanova-2008, Romanova-2021, Cemeljic-2019}.

Such a short time-scale does not allow for a similar explanation of the smooth, wave-like brightness variations observed in some CTTS, which have characteristic time-scales of $\gtrsim 10$\,yr. The existence of these variations has long been known from photographic \citep{Kholopov-1970} and photoelectric \citep{Grankin-07} observations. In the case of binary CTTS, long-period brightness variations can be attributed to modulation of the accretion rate due to tidal interactions of the companion with the disc of the primary component — see, for example, \citet{Lamzin-2001}. The cause of such phenomena in single CTTS remains unresolved.

In this context, we turned our attention to BP~Tau, whose variability was 
first discovered by I.~Lehmann-Balanowskaja 90 years ago 
\citep{Lehman-Balanowskaja-1935}. BP~Tau is a single CTTS \citep{Kounkel-2019} 
actively accreting matter ($\dot M_{\text ac} \sim 0.2-1\times 
10^{-8}$\,M$_\odot$\,yr$^{-1}$ \citealt{Long-2011, Nisini-2024}) from 
its protoplanetary disc, which was detected using the ALMA 
interferometer \citep{Long-2019}. Observations reveal material outflows 
originating from the vicinity of the star \citep{Hartigan-1995, 
Alencar-2000, Errico-2001}, with part of the ejected material 
being collimated into a bipolar jet \citep{Dodin-2024}. According to  
\citet{Herzeg-Hillebrandt-2014}, the effective temperature, 
extinction, and luminosity of BP~Tau are $T_{\text eff}=3770\pm 150$\,K, 
(spectral type M0.5), $A_{\text V}=0.45$ and $0.4$\,$L_\odot$, respectively, 
at a distance of $d\approx 130$\,pc \citep{Akeson-2019}.

\citet{Donati-2008} concluded that the magnetic field of BP~Tau can be represented as the sum of a dipole ($B=1.2$\,kG) and octupole ($B=1.6$\,kG), with their axes inclined at angles of $20\degr$ and $10\degr$, respectively, relative to the rotation axis of the star, such as the hot spot covers several percent of the stellar surface and is elongated along a meridian. According to calculations by \citet{Long-2011}, in this configuration, the interaction between the disc and the stellar magnetosphere should result in the localisation and parameters of the hot spot remaining largely unchanged over time. However, no stable photometric period has been detected for BP~Tau: it appears that the period varies by several percent across different observing seasons, fluctuating around a value of $\approx 8^{\text d}$ \citep{Rebull-2020, Lin-2023, Wendeborn-II-2024}. Furthermore, the inclination of the rotation axis of the disc -- and thus the star -- relative to the line of sight is relatively large: $i=38.1\pm 0.5\degr$ \citep{Long-2019}. \citet{Kiryukhina-2024} also found no regular periodicity in the variability of the star’s emission line intensities.

This likely implies that the parameters of the disc and/or the configuration of the star's magnetic field are not constant over time. In this context, it is appropriate to examine how the photometric behaviour of BP~Tau has evolved over a long-term interval, which is the focus of this study.

The paper is organised as follows. In Section~2, we describe how the observational data underpinning our study were obtained. In Section~3, we present the results derived from these data. In Section~4, we propose an interpretation of the findings. The main outcomes of the study are summarised in Section~5.

                 
\section{Observational data}
 \label{sect:observation}

To construct the historical light curve of BP~Tau, in addition to data from the literature (see Section 3.3.1), we visually estimated the star's brightness using 458 photographic plates from the SAI MSU collection. These plates were obtained between November 27, 1911 and February 15, 1988 in a photometric system close to the Johnson $B$-band. The accuracy of the brightness estimates was no worse than $\approx 0\fm 2.$ The results are presented in Table 1, where, as throughout this work, we use the reduced Julian date ${\text rJD}={\text JD}-$2,400,000.

We also obtained photometry of~BP Tau in the visible range using the 60-cm telescope at the Caucasian Mountain Observatory (CMO) of SAI MSU, equipped with a CCD camera and a set of $BVRI$ filters from the Bessel-Cousins system \citep{Berdnikov2020}. The brightness of the comparison stars was taken from the AAVSO website. \footnote{https://www.aavso.org}

\begin{table}
\renewcommand{\tabcolsep}{0.2cm}
 \caption{Photographic magnitudes}
  \label{tab:B-lc}
  \begin{tabular}{ll|ll}
\hline
rJD & $m_{\text pg}$  & rJD & $m_{\text pg}$ \\
  \hline
19\,368.410 & 13.1 & ........... & .... \\
........... & .... & 47\,207.237 & 13.3 \\
  \hline
 \end{tabular} \\
\end{table}

In December 2023, polarimetric observations of BP~Tau were conducted in the $V,$ $R,$ and $I$-bands using the speckle polarimeter SPP \citep{Safonov-17} on the 2.5-m telescope at the CMO. Details of the observation process and data reduction are described in \citet{Dodin-19}. The results, including the degree and angle of polarisation along with their uncertainties for each band, are presented in Table~\ref{tab:BP-polariz}.

\begin{table*}
\renewcommand{\tabcolsep}{0.2cm}
 \caption{Results of polarimetric observations of BP~Tau in the $V,$ $R,$ $I$ bands.}
  \label{tab:BP-polariz}
  \begin{tabular}{ll|llll|llll|llll}
\hline
Date & UT & $p_V$ & $\sigma_p$ & $\theta_V$ & $\sigma_\theta$ &
$p_R$ & $\sigma_p$ & $\theta_R$ & $\sigma_\theta$ &
$p_I$ & $\sigma_p$ & $\theta_I$ & $\sigma_\theta$ \\
\hline
07.12.2023 & 18:14 &      &      &       &      &  0.58 & 0.24 & 38.7  
& 23.6 & 0.89 & 0.17 & 59.2 & 11.2 
\\
19.12.2023 & 19:48 &  0.69 & 0.17 & 84.0  & 13.9 & 0.34 & 0.19 & 50.3 & 30.8 
& 0.56 & 0.16 & 72.4  & 16.7 \\
29.12.2023 & 21:56 &  0.44 & 0.16 & 134.0 & 21.1 & 0.12 & 0.16 & 75.7 & 76.3 
& 0.10 & 0.15 & 119.9 & 92.0 \\
  \hline
 \end{tabular} \\
\end{table*}

To obtain quantitative constraints on the presence of a close companion around BP~Tau, we conducted observations using speckle interferometry. On December 19, 2023 at 19:44 UT, we acquired 9\,791 frames in the $I$-band, each with an exposure time of 22.9\,ms. The half-width of the long-exposure image was $1.12^{\prime\prime}$.

With the same goal in mind, we investigated the radial velocity variability of BP~Tau using 45 archival spectra of the star obtained with the ESPaDOnS spectrograph (CFHT) over four observational seasons from February 2006 to January 2014, as well as five spectra obtained with the ESPRESSO spectrograph (VLT) in December 2022 \citep{Kiryukhina-2024}. To determine the radial velocity of BP~Tau in each individual spectrum, we calculated the weighted mean of the velocities of 250 absorption lines with minimal blending. The absorption spectrum of the WTTS TAP~45, which has similar stellar atmosphere parameters, was used as the reference spectrum against which the line velocities were computed using the least-squares method \citep{Kiryukhina-2024}. The resulting accuracy of the radial velocity determination for individual spectra is $\approx 10$\,m\,s$^{-1}$.

                
\section{Results}
\subsection{Historical light curve of BP~Tau}  
\label{sec:lcurve}

\begin{figure*}
    \centering
    \includegraphics[width=0.95\textwidth]{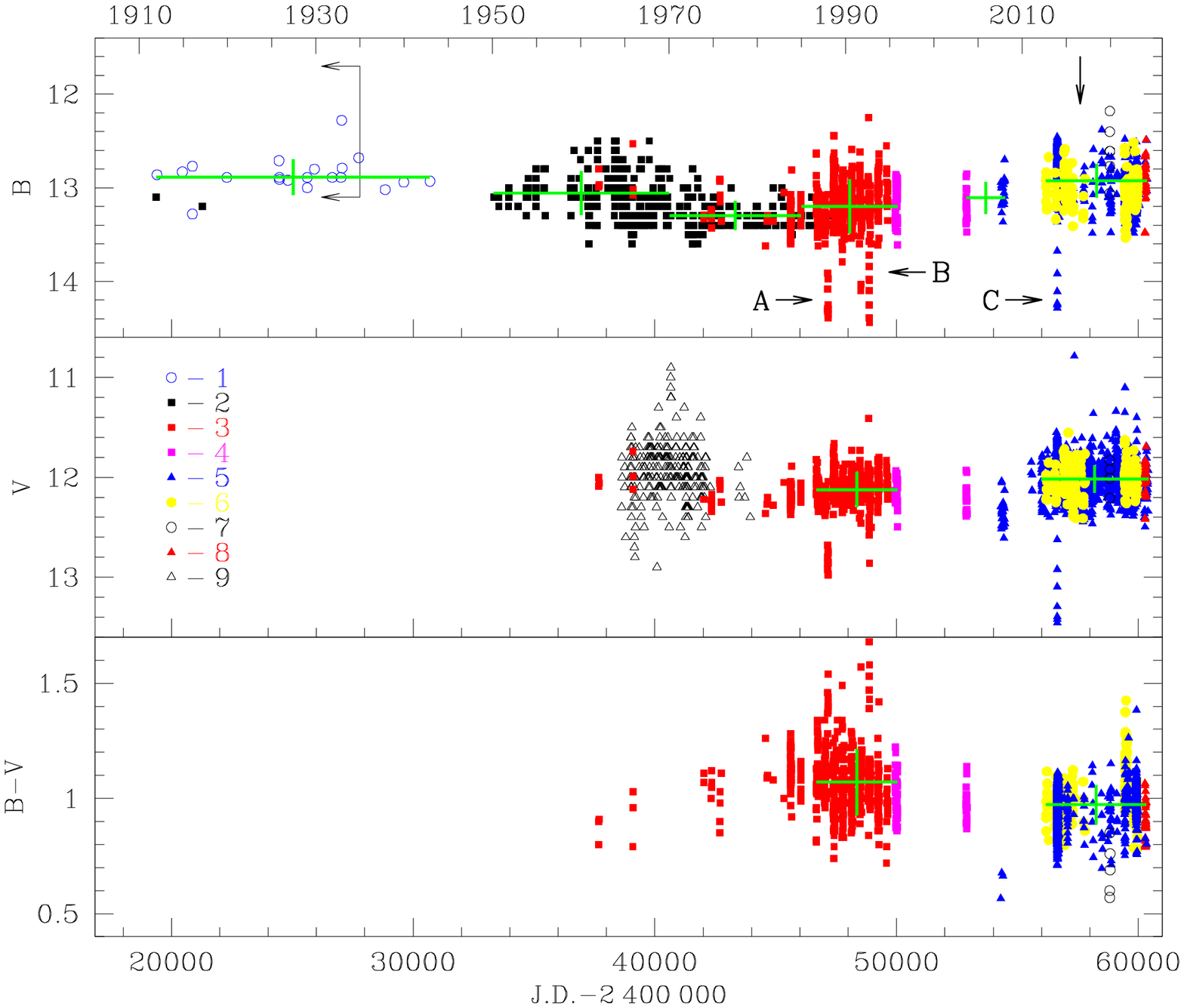}
    \caption{Secular light curve of BP~Tau in the $B$ and $V$-bands and the corresponding variation in the colour index $B-V$. The data are represented by various symbols and colours, sourced from the following works: 1 -- \citet{Kholopov-1951}, 2 -- our estimates from photographic plates, 3 -- \citet{Herbst-94}, 4 -- \citet{Grankin-07}, 5 -- AAVSO, 6 -- our photoelectric and CCD data (CrAO), 7 -- \citet{Robinson-2022}, 8 -- CCD observations (CMO), 9 -- visual data from AAVSO. Green crosses indicate the mean values with their uncertainties within the corresponding time intervals. Horizontal arrows labelled A, B, and C mark episodes of deep brightness dimming, which are shown in greater detail in Figures~\ref{fig:AB-dip} and \ref{fig:C-dipS}. A vertical arrow indicates the approximate moment of microjet emergence, as reported by \citet{Dodin-2024}.
    }
    \label{fig:bp-lcurve}
\end{figure*}

To construct the historical light curve of BP~Tau, we supplemented our measurements  (Sect.~\ref{sect:observation}) with data from the papers by \citet{Kholopov-1951}, \citet{Robinson-2022}, \citet{Grankin-07}, as well as from the databases of \citet{Herbst-94} and AAVSO. We also used unpublished results from our photoelectric (September 2012 – October 2015) and CCD (September 2014 – January 2023) observations, conducted at the Crimean Astrophysical Observatory (CrAO) using the 1.25-m AZT-11 telescope equipped with a photometer-polarimeter \citep{Piirola75}, as well as CCD cameras FLI ProLine PL230 (2015–2021) and greateyes ELSEi (2022–2023). The resulting light curve is shown in Fig.~\ref{fig:bp-lcurve}. The symbol $']'$ in the figure reflects the fact that in the publication by \citet{Lehman-Balanowskaja-1935}, only the range of brightness variations is provided, without specifying the time interval to which this range corresponds.

\begin{figure*}
    \includegraphics[width=0.8\textwidth]{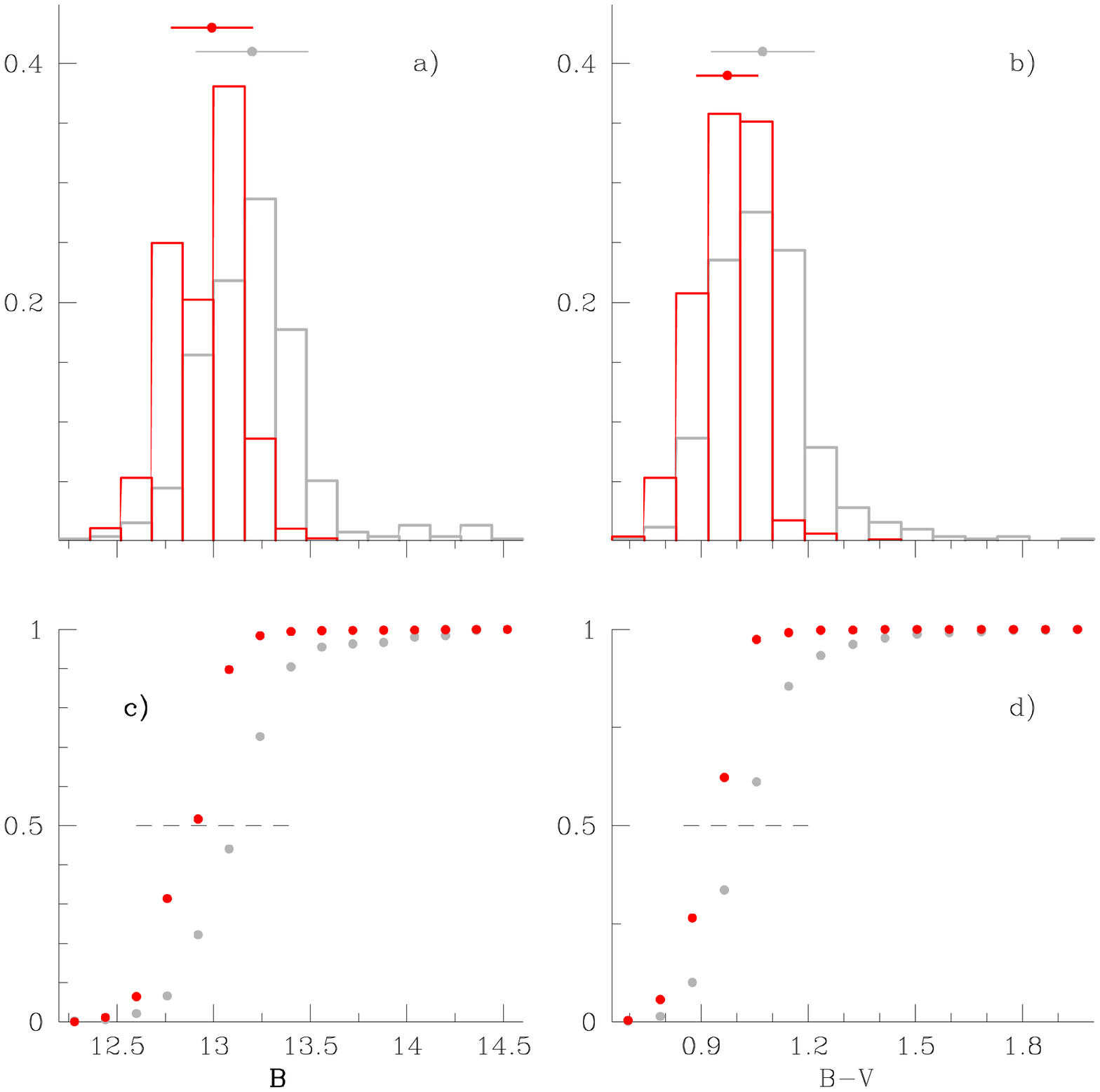}
    \caption{Histograms depicting the brightness variations in the $B$-band (a) and the colour index $B-V$ (b) during the seasons 1985–1997 (grey, $\approx 500$ measurements) and 2010–2024 (red, $\approx 2000$ measurements). Coloured circles and lines above the histograms indicate the mean values of $B$ and $B-V$, along with their corresponding root-mean-square deviations, for the respective time intervals. Panels (c) and (d) show the cumulative distribution functions for each of the above histograms. The dashed lines determine the median values of the respective quantities.
    }
    \label{fig:gistogramm}
\end{figure*}

It can be seen from the upper panel of the figure, that BP~Tau exhibits strong short-term brightness fluctuations in the $B$-band on time-scales of 1-10\,days \citep{Gotz-1961, Robinson-2022, Wendeborn-II-2024}, while also undergoing a gradual variation in its average brightness over the past 100 years. Specifically, the average brightness gradually decreased until around 1980, after which it began to rise steadily and has now returned to a level similar to that observed a century ago. For example, between 1970 and 1985, the star was, on average, $0\fm 38$ fainter in the $B$-band than during the period from 2010 to 2024, a difference that is 2.2 times greater than the characteristic $\sigma_{\text B}$ for these decades. Furthermore, as shown in Fig.~\ref{fig:gistogramm}a, the histograms of brightness distributions for these two epochs, normalised to the total number of observations in each epoch, appear different, as do the corresponding cumulative probability distribution functions (Fig.~\ref{fig:gistogramm}c). Using the Kolmogorov-Smirnov test \citep{matstat2}, we found that the probability of this difference being due to random causes is less than 0.1\,\%{}. Based on this, we conclude that the nature of the brightness variations of BP~Tau in the $B$-band during the two considered epochs was indeed different.  

It appears that the mean $V$ brightness also reached a minimum around 1980 (see the middle panel of Fig.~\ref{fig:bp-lcurve}). However, we cannot state this with absolute certainty, as most of the data from the earlier epoch consist of visual brightness estimates, which may have relatively large uncertainties. With similar confidence, we can assert that after 1980, the mean colour index $B-V$ began to decrease (lower panel of Fig.~\ref{fig:bp-lcurve}). Formally, the difference in the mean values of $B-V$ between the periods 1985–1997 and 2010–2024 is relatively small: $1.07 \pm 0.14$ and $0.97 \pm 0.09$, respectively. However, as shown in Fig.~\ref{fig:gistogramm}b, the histograms of $B-V$ distributions and their corresponding cumulative probability distribution functions (Fig.~\ref{fig:gistogramm}d) differ noticeably between these two epochs. The Kolmogorov-Smirnov test indicates that, with a probability greater than 99.9\,\%{}, the difference in the behaviour of the $B-V$ colour index is not due to statistical error, i.e., it is real.

The arguments presented above provide grounds to assert that the mean brightness of BP~Tau in the $B$ and $V$-bands undergoes long-term variations with an amplitude of $\Delta B\approx 0\fm 2$ and a characteristic time-scale of several decades. Given the star's effective temperature $T_{\text eff} \lesssim 4000$\,K \citep{Herzeg-Hillebrandt-2014}, the $B-V$ colour index of its photosphere should exceed 1.7, even without accounting for interstellar reddening \citep{Pecaut-Mamajek-2013}. As can be seen from Fig.~\ref{fig:bp-lcurve}, the observed $B-V$ value is less than this threshold, indicating that we are always observing a hot spot. Consequently, the secular variability in the brightness of BP~Tau is likely associated either with changes in the spot parameters or with variable extinction of its emission.


 \subsection{Deep dimming events of BP~Tau}
 \label{sec:aflare}

\begin{figure*}
    \centering
    \includegraphics[width=0.95\textwidth]{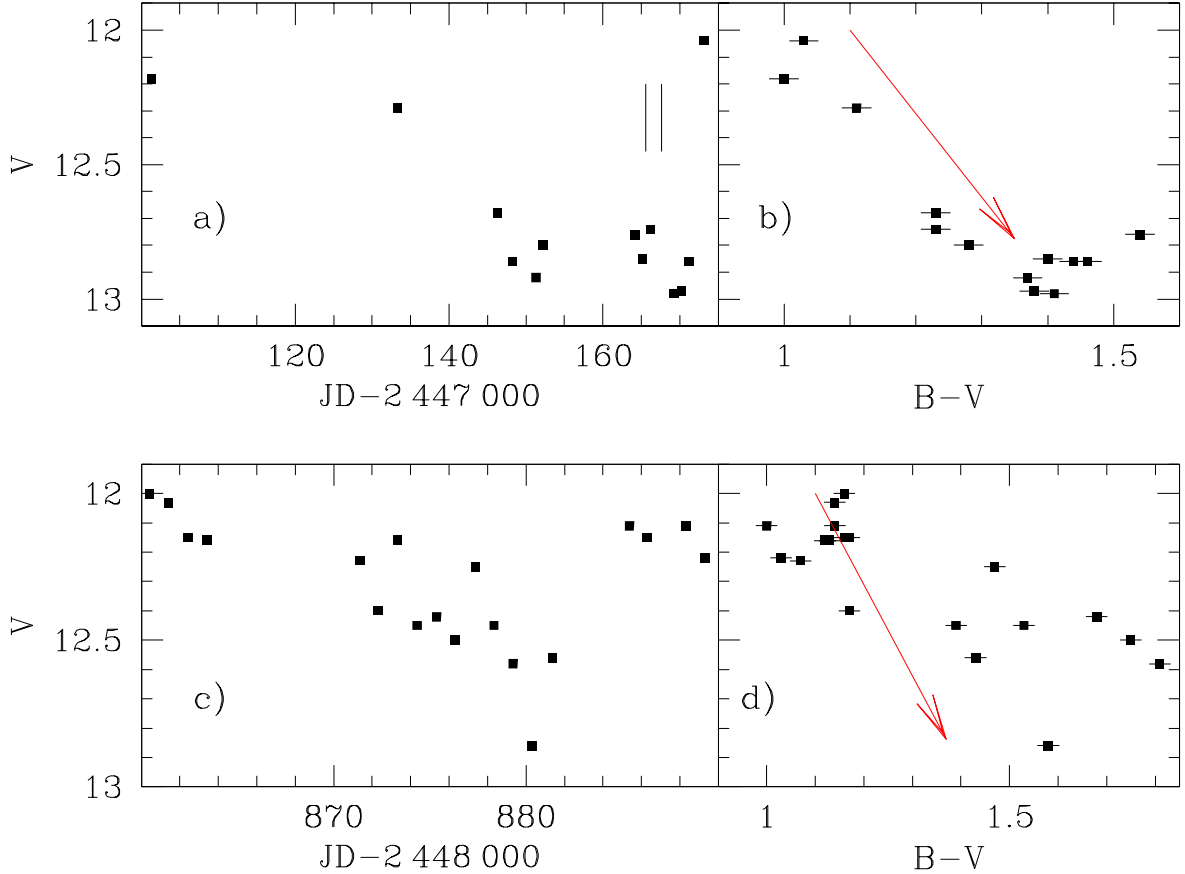}
    \caption{Variations in the $V$-band brightness and the $B-V$ colour index around the deep dimming events of~BP Tau: the upper and lower rows correspond to episodes A and B in Fig.~\ref{fig:bp-lcurve}, respectively. Vertical lines in panel (a) mark the times at which the spectra analysed by \citet{Hartigan-1995} were obtained. The arrow in the $V$ vs. $B-V$ diagrams indicates the direction of reddening according to the standard extinction law. See text for details.}
    \label{fig:AB-dip}
\end{figure*}

We identified three episodes of significant dimming $(\Delta B \approx 1\fm 5)$ of the star, labelled A, B, and C in Fig.~\ref{fig:bp-lcurve}. Detailed light variations in the $V$-band for episodes A (late December 1987) and B (mid-September 1992) are presented in Figs.~\ref{fig:AB-dip}a and \ref{fig:AB-dip}c, respectively.

From Figs.~\ref{fig:AB-dip}b and \ref{fig:AB-dip}d, it is evident that the dimming during these episodes was accompanied by an increase in the $B-V$ colour index, indicating that the star became redder. Qualitatively, this $V$ vs. $B-V$ relationship can be naturally explained by a decrease in the contribution from the hot spot emission. However, the underlying cause remains unclear: whether it results from a decrease in the accretion rate or the shielding of the hot spot's radiation.

\begin{figure}
    \centering
    \includegraphics[width=0.45\textwidth]{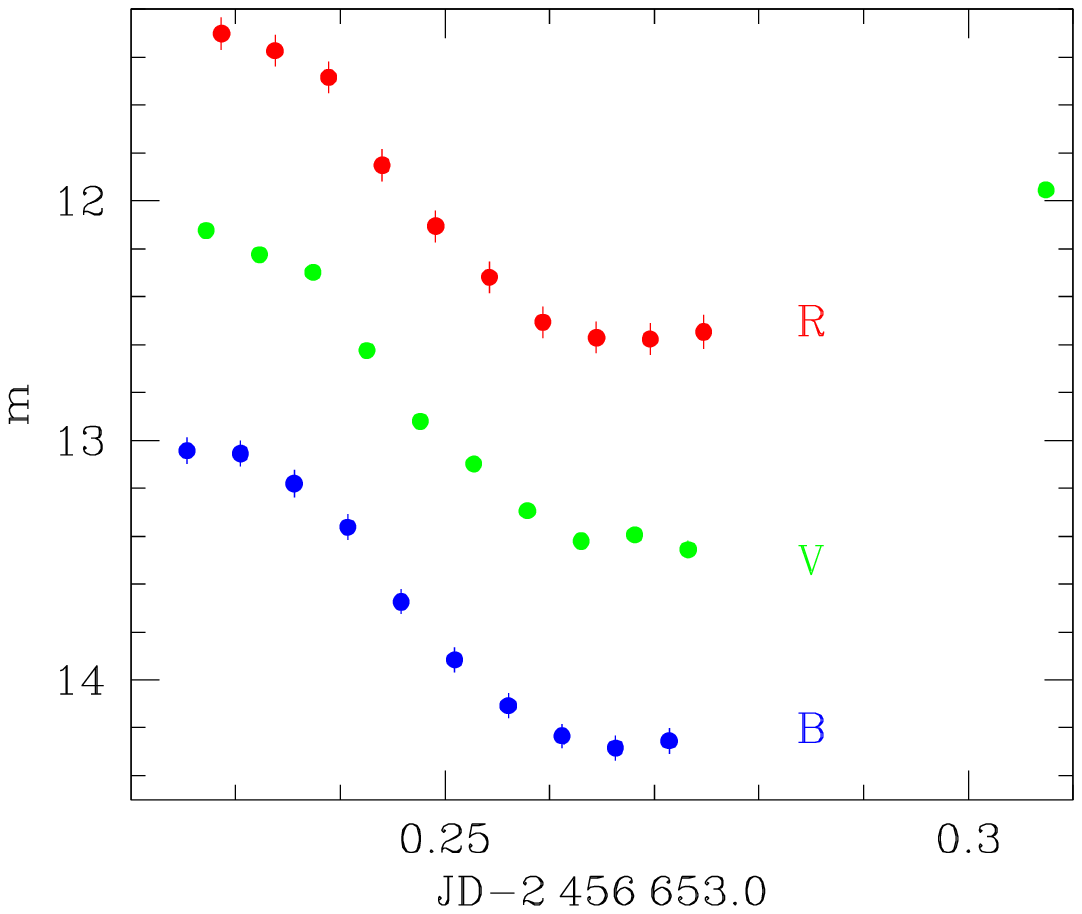}
    \caption{Variations in brightness in the $B$, $V$ and $R$-bands on December 26, 2013 (episode C in Fig.~\ref{fig:bp-lcurve})}
    \label{fig:C-dipS}
\end{figure}

A distinctly different pattern was observed on December 26, 2013 (episode C) -- see Fig.~\ref{fig:C-dipS}. In this case, the dimming in the $V$-band was approximately $0\fm5$ deeper than in the previous episodes, with both the decline and recovery to the original level occurring much more rapidly, within just 1.5~hour. Furthermore, the brightness drop occurred while the colour indices remained nearly constant (within the error limits): $B-V\approx 0.83$ and $V-R\approx0.86.$  

\begin{figure*}
    \centering
    \includegraphics[width=0.95\textwidth]{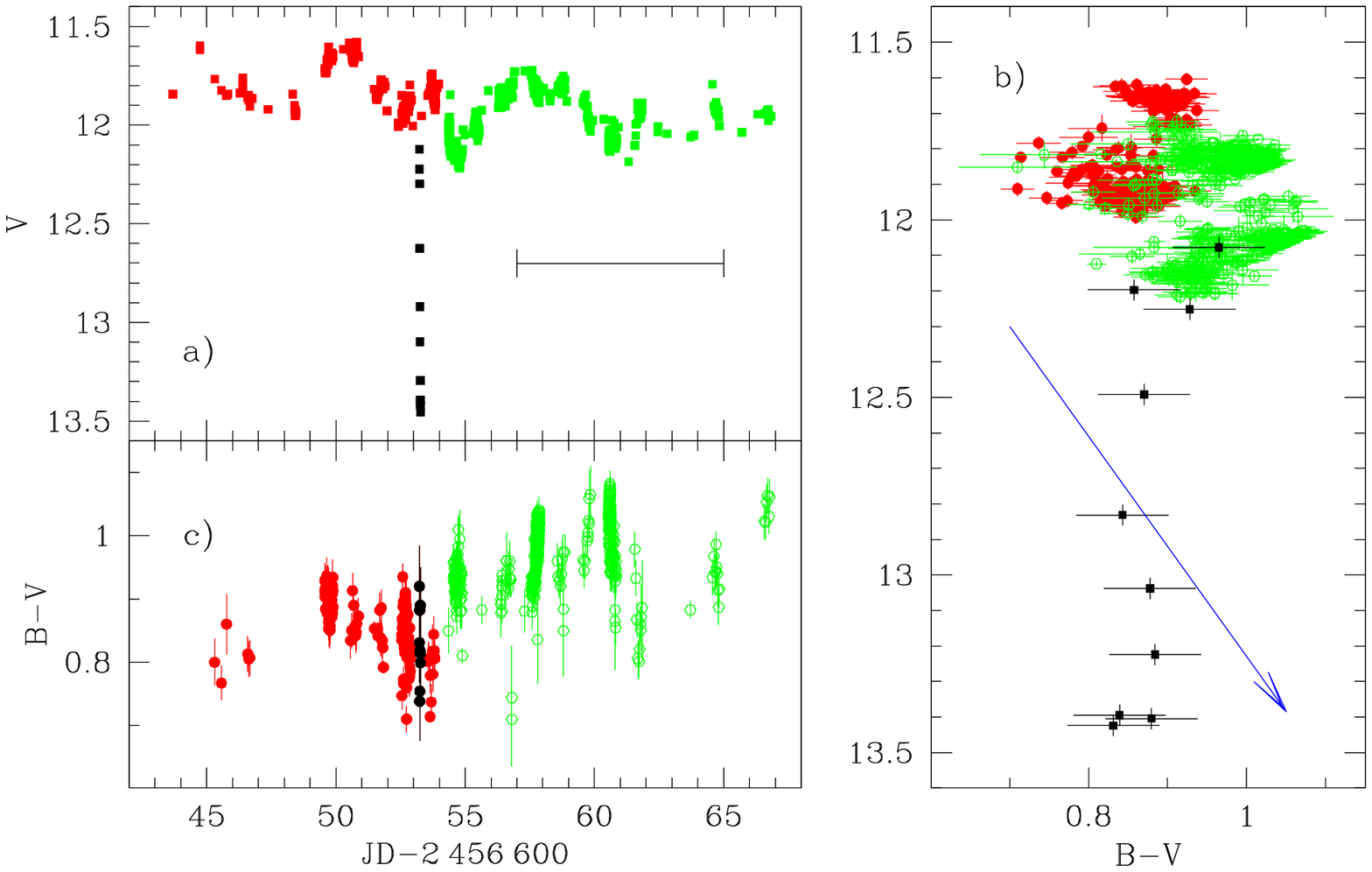}
    \caption{Left panels: Variations in the $V$-band brightness and $B-V$ colour index around December 26, 2013 (episode C in Fig.\ref{fig:bp-lcurve}). Right panel: The colour-magnitude diagram ($V$ vs. $B-V$) for this segment of the light curve. The arrow in the $V$ vs. $B-V$ diagram shows the direction of reddening according to the standard extinction law. The horizontal segment corresponds to the approximate axial rotation period ($P=8^{\text{d}}$). Different parts of the light curve, characterised by distinct $V$ vs. $B-V$ relationships, are highlighted in different colours. Further details are provided in the text.}
    \label{fig:C-dipL}
\end{figure*}

Approximately 6~h after this episode, the previously observed decrease in the $B-V$ colour index gave way to an increase -- see Figs.~\ref{fig:C-dipL}a and \ref{fig:C-dipL}c. In other words, the star became bluer before the sharp dimming event and began to redden afterward -- see also Fig.~\ref{fig:C-dipL}b. This leads us to conclude that the extremely short and deep dimming event, recorded by only one observer (A. Rodda, AAVSO), is indeed a real phenomenon. The dimming occurred with nearly constant (and very small) values of the $B-V$ colour index, suggesting that it cannot be explained by a reduction in the brightness of the hot spot due to a decrease in the accretion rate. Possible causes of the deep dimming events of BP~Tau will be discussed in Section~\ref{sec:discuss}.

\subsection{Constraints on the parameters of a hypothetical satellite}
\label{sec:sputnik}

\citet{Kounkel-2019} found no evidence of radial velocity variability in BP~Tau from infrared spectra obtained with the APOGEE spectrograph on the 2.5-m telescope at Apache Point Observatory. Based on this result, they concluded that the star does not have a companion. However, their study did not include quantitative constraints on the parameters of a potential companion, prompting us to address this shortcoming. To do so, we analysed 45 archival spectra of the star (Sect.~\ref{sect:observation}) to investigate how its radial velocity $V_{\text r}$ changes over time.

We fitted the obtained radial velocity values with a periodic function, testing trial periods $P$ in the range from 50 to 40,000 days:
\begin{equation}
    V_{\text r}\cos({\varphi + \Delta\varphi}) + \bar{v}, \quad \varphi = \frac{2\pi t}{P}.
\end{equation}
The parameters of the sinusoid $V_{\text r},$ $\bar{v},$ $\Delta\varphi$ and their uncertainties were determined using the least-squares method.

The protoplanetary disc of BP~Tau appears to be axially symmetric \citep{Jennings-2022}, with both its inner and outer regions inclined to the line of sight at the same angle \citep{Dodin-2024}. This suggests that the orbit of a hypothetical companion lies in the plane of the disc, with an eccentricity $e=0$. In this case, the inclination of the orbit to the line of sight should match that of the disc, i.e., $i=38\fdg 1 \pm 0\fdg 5$ \citep{Long-2019}.  Consequently, the orbital velocity of the primary star $V$ is related to its radial velocity $V_{\text r}$ by the equation $V=V_{\text r} /\sin i$.

The mass function of the brighter star in a binary system (assuming $e=0$) is given by:
\begin{equation}
\frac{m^3 \, {\sin{i}}^3}{(M + m)^2} = 1.038\times 10^{-7} V_{\text r}^3 P\, ,
    \label{eq:f-mass}
\end{equation}
where $V_{\text r}$ is the amplitude of the radial velocity variation of the brighter star in km\,s$^{-1}$, $P$ is the period in days, $M \approx 0.5 M_{\odot}$ is the mass of BP~Tau (see below), $m$ is the mass of the companion in $M_{\odot}$, and $i$ is the inclination of the orbit to the line of sight.

\begin{figure}
    \includegraphics[width=0.5\textwidth]{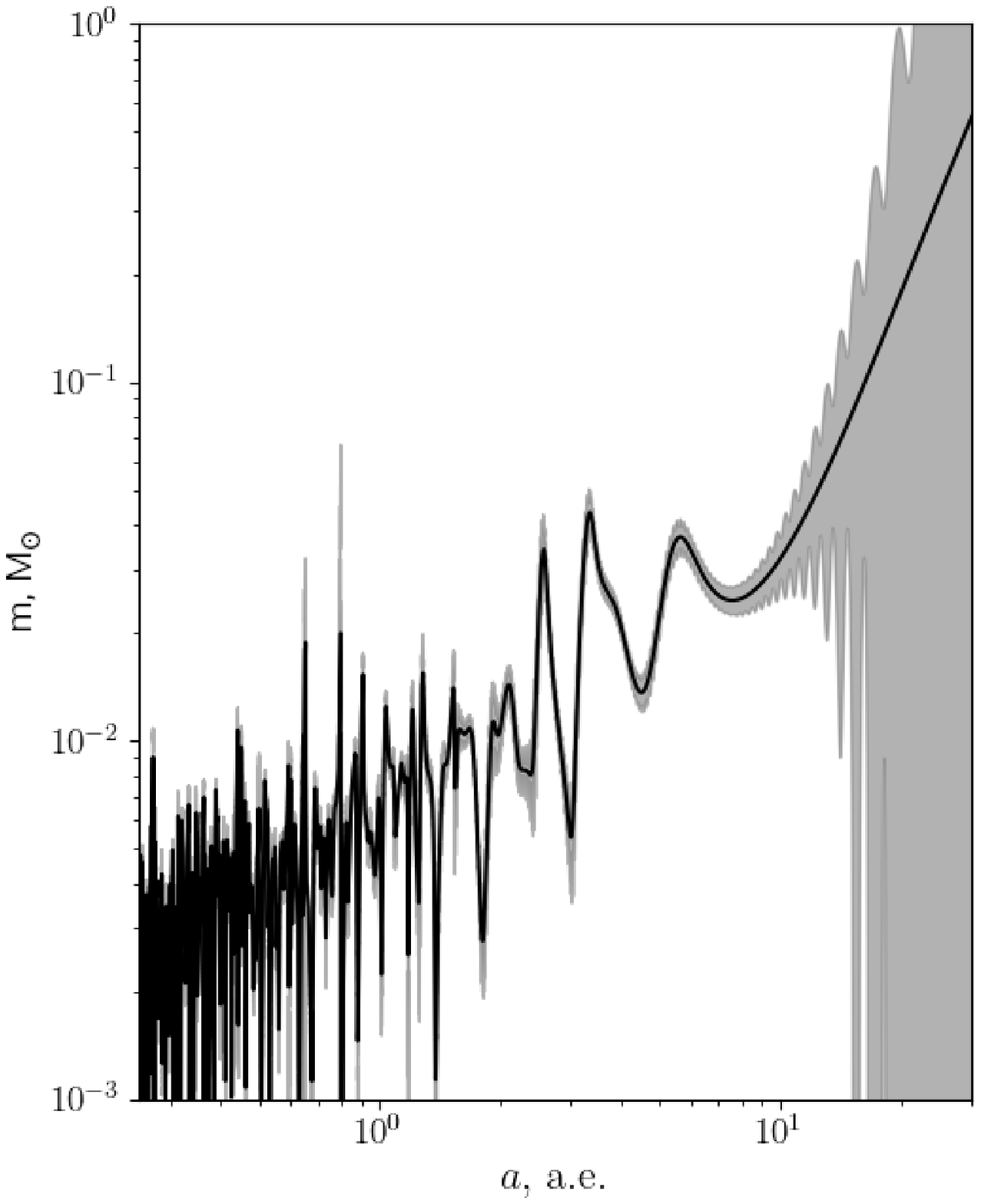}
    \caption{Upper limit on the mass of a hypothetical companion to BP~Tau as a function of stellar separation, derived from radial velocity measurements.}
    \label{fig:M-vs-a}
\end{figure}

It then follows that
\begin{equation}
    m = 1.038\times 10^{-7} V_{\text r}^3 P \left(1 + 
\frac{M}{m}\right)^2 \frac{1}{{\sin{i}}^3}\, .
 \label{eq:fm2}
\end{equation}
By numerically solving this equation for $m$, we obtain the possible mass of the companion for a given period value. The semi-major axis $a$ is then calculated using Kepler's third law:
\begin{equation}
    a = {\left( M+m \right)}^{1/3}\, P^{2/3}.
     \label{eq:kepler}
\end{equation}
The upper limit of the companion's mass $m$, as derived from this relation and depending on its distance $a$ from BP~Tau (assuming $M=0.5$~M$_\odot$), is shown in Fig.~\ref{fig:M-vs-a}.

\begin{figure}
    \includegraphics[width=0.50\textwidth]{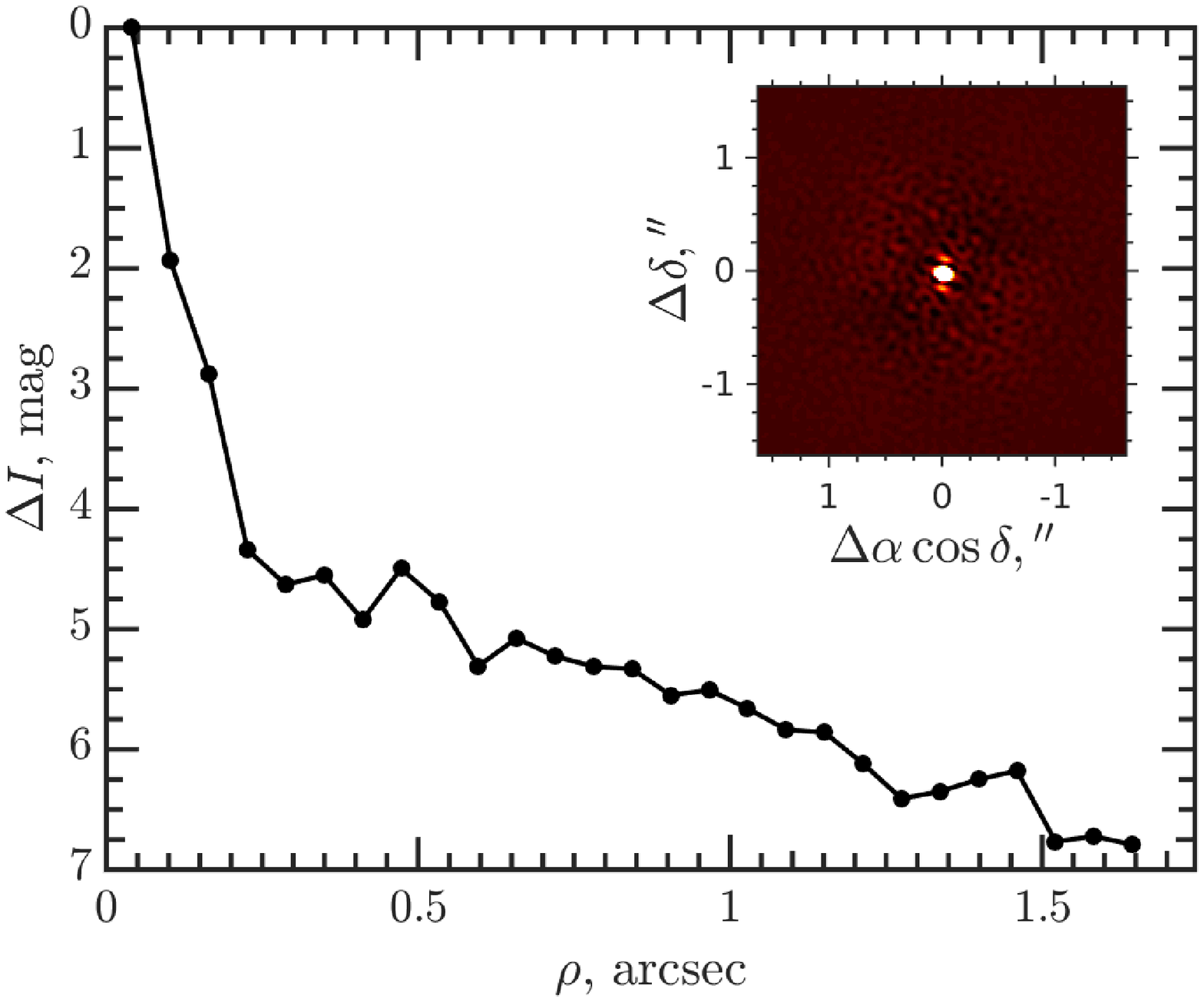}
    \caption{Constraints on the brightness difference between BP~Tau and its hypothetical companion in the $I$-band, as a function of their separation, derived from speckle-interferometric data. The inset displays the autocorrelation function.}
    \label{fig:SPP} 
\end{figure}

Another approach to estimating the parameters of a potential companion to BP~Tau relies on our speckle-interferometric observations (Sect.~\ref{sect:observation}), which found no evidence of binarity. Using the autocorrelation function derived from these observations, we applied the methodology outlined in \citet{Strakhov-2023} to determine the upper limit on the separation between the star and its companion as a function of the brightness difference between the components in the $I$-band. The results are presented in Fig.~\ref{fig:SPP}.

\begin{figure*}
    \centering
    \includegraphics[width=0.85\textwidth]{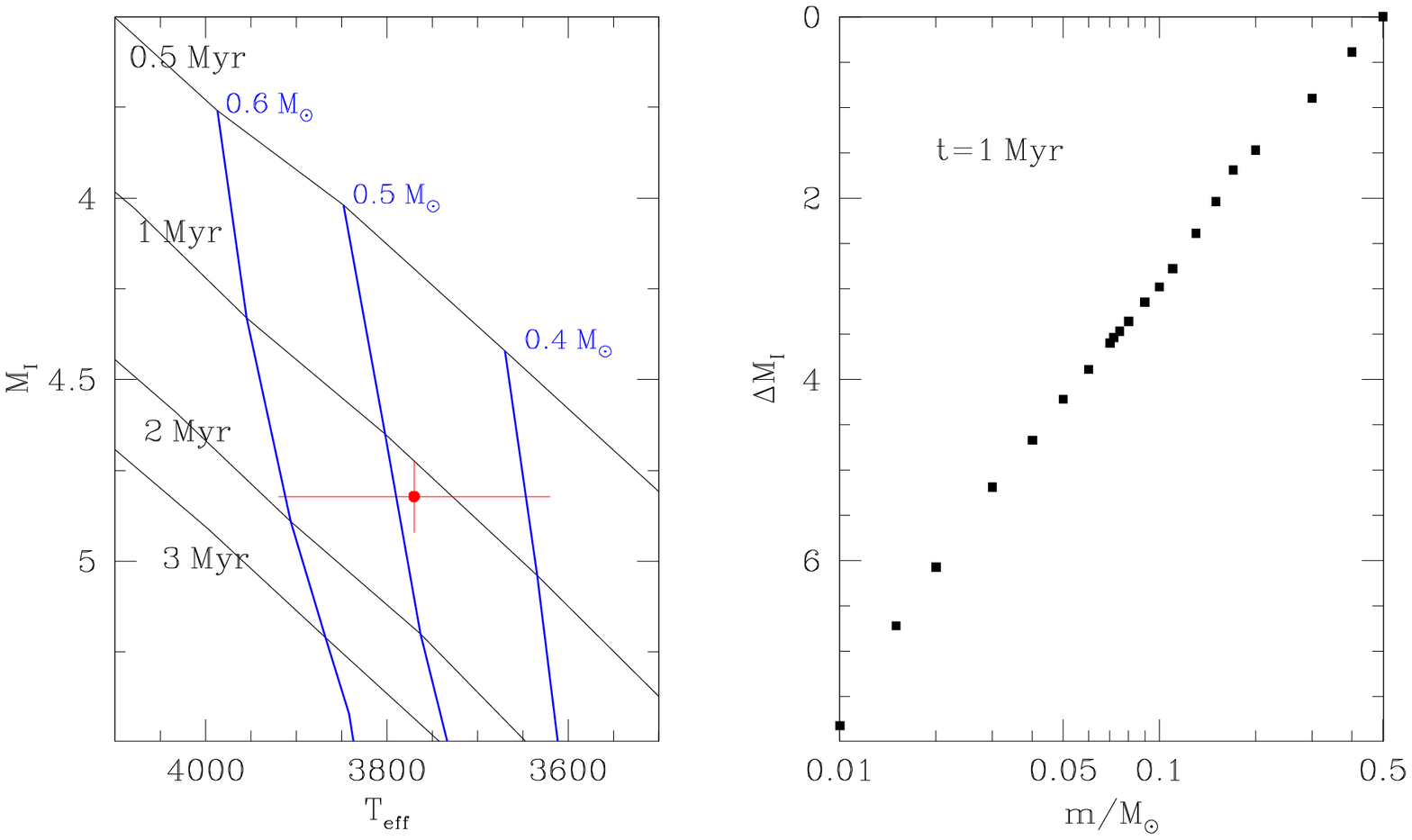}
    \caption{The position of BP~Tau on the effective temperature -- absolute $I$ magnitude diagram (left panel), and the difference in the $I$-band brightness between BP~Tau $(M=0.5$\,M$_\odot)$ and stars of mass $m$ at an age of 1\,Myr (right panel). The plots are based on data from \citet{Baraffe+2015}.}
    \label{fig:age}    
\end{figure*}

To compare the results of speckle-interferometric observations of~BP Tau in the $I$ band with the above estimate of the $M \div a$ dependence derived from radial velocities, we proceeded as follows. First, we determined the absolute magnitude $M_{\text I}$ of BP~Tau in the $I$-band using its brightness in this range, measured from our photometric observations: $m_{\text I}=10\fm 33 \pm 0\fm 09.$ For this calculation, we assumed a distance to the star of $d=130$~pc \citep{Akeson-2019} and an extinction $A_{\text I}=0.27,$ corresponding to $A_{\text V}=0.45$ \citep{Herzeg-Hillebrandt-2014} with $R_{\text V}=3.1.$ Assuming an effective temperature of $T_{\text eff}=3770 \pm 150$\,K \citep{Herzeg-Hillebrandt-2014}, and using theoretical evolutionary tracks and isochrones for young stars from \citet{Baraffe+2015}, we found that the mass of BP~Tau is $\approx 0.5$\,M$_\odot,$ with an age close to 1\,Myr -- see the left panel of Fig.~\ref{fig:age}. We note that the mass of BP~Tau, as determined from ALMA interferometric observations \citep{Long-2019}, was found to be $0.52_{-0.12}^{+0.15}$\,M$_\odot.$

For stars on the 1-Myr isochrone, we constructed the dependence $\Delta M_{\text I} (m)$, which describes how the $I$-band brightness of a star of mass $m$ differs from that of a $0.5$\,M$_\odot$ star -- see the right panel of Fig.~\ref{fig:age}. Since $\Delta M_{\text I}=\Delta m_{\text I}$, this relationship allows us to convert the observed value of $\Delta m_{\text I}$ into a mass difference between the components. Furthermore, the apparent separation between the components, $\rho$, is related to the radius of the (circular) orbit by the relation $\rho=a\cos i,$ when converting angular measurements into linear ones.

\begin{figure}
    \includegraphics[width=0.45\textwidth]{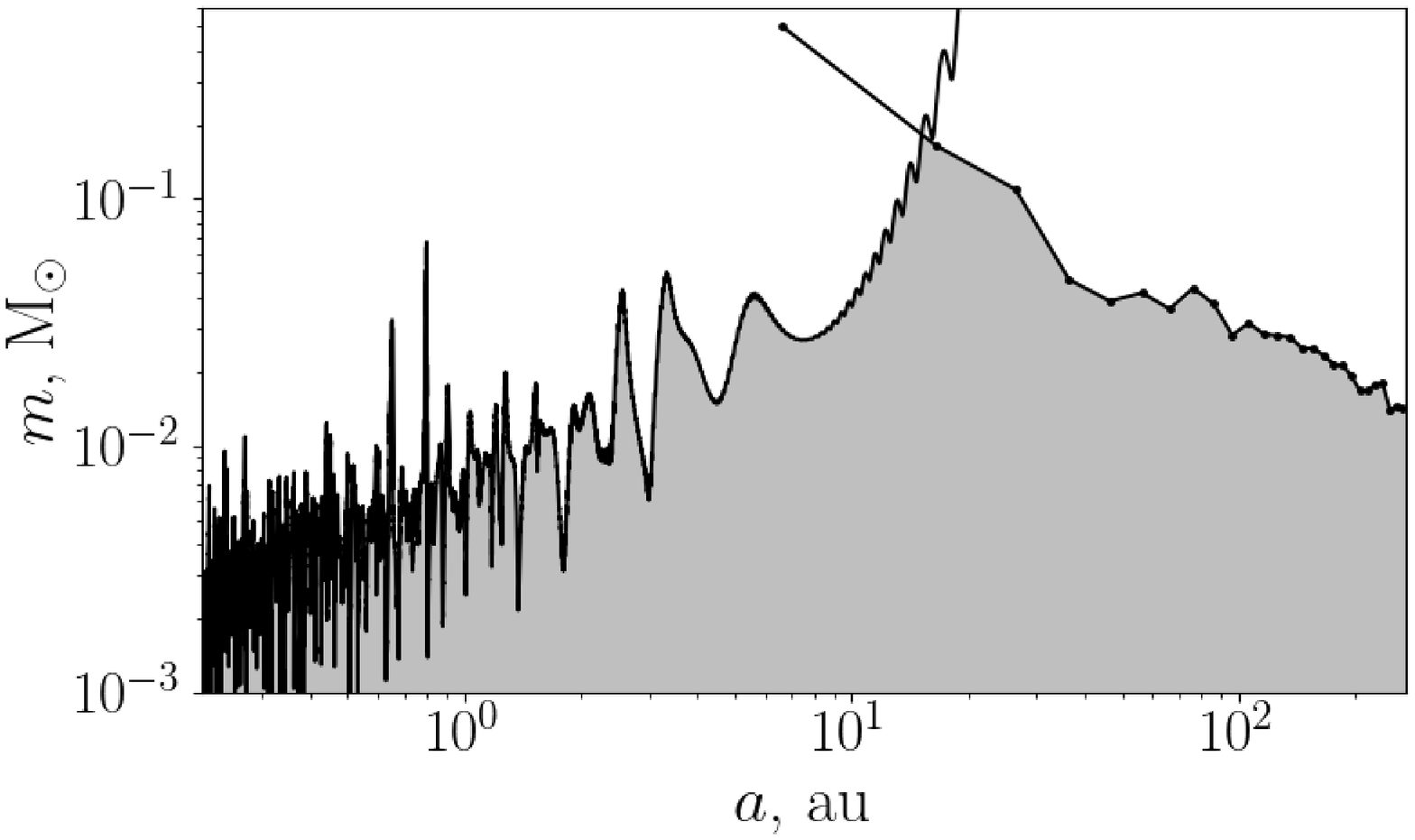}
    \caption{Constraints on the mass of a hypothetical companion to BP~Tau as a function of its separation from the primary star, obtained by combining the data from Figs.~\ref{fig:SPP} and \ref{fig:M-vs-a}. The shaded region corresponds to the allowed parameters of the companion.}
    \label{fig:BP_m2} 
\end{figure}

The combined results of speckle-interferometric and spectroscopic observations of BP~Tau are presented in Fig.~\ref{fig:BP_m2}. From the figure, it follows that if BP~Tau has a companion, then within the distance ranges of 0.1–15\,AU and 30–200\,AU, the companion is likely a brown dwarf or planet $(m<0.1$\,M$_\odot)$. In the range of 15–30\,AU, the mass of the hypothetical companion does not exceed  $\approx 0.2$\,M$_\odot.$

\section{Discussion}
\label{sec:discuss}

\begin{figure*}
    \centering
    \includegraphics[width=0.95\textwidth]{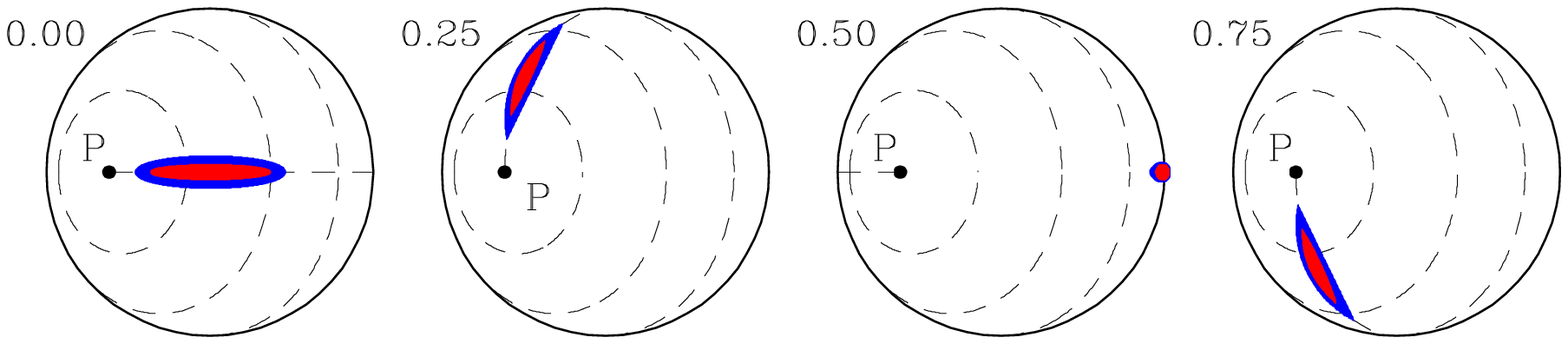}
    \caption{A semi-qualitative depiction of the spot's motion relative to the observer, based on data from \citet{Long-2011}. The positions of the spot are shown for rotational phases $\varphi = 0.00,$ 0.25, 0.50, and 0.75. The point $P$ represents the star's pole, with dashed lines indicating parallels at $30\degr,$ $60\degr$, and $90\degr$, as well as the meridian passing through the centre of the spot. The red region highlights the area where the majority of the spot's bolometric flux is emitted -- see the text.}
    \label{fig:scheme} 
\end{figure*}

As noted in Sect.~3.1, the colour index $B-V$ of BP~Tau is consistently much smaller than the value expected for its effective temperature. This likely indicates the presence of a hot (accretion) spot on the stellar surface, which is always visible from the Earth. As mentioned in the Introduction, \citet{Donati-2008} concluded that the hot spot covers a few percent of BP~Tau's surface and is elongated along the meridian -- a result confirmed by three-dimensional MHD simulations conducted by \citet{Long-2011}. Fig.~\ref{fig:scheme} illustrates the motion of such a geometrically shaped spot relative to an observer as the star rotates about its axis. For clarity, we represented the actual bolometric intensity distribution within the hot spot, as shown in Fig.~9 of \citet{Long-2011}, with an oval-shaped spot, highlighting an inner region that emits over 80\,\%{} of the spot's bolometric flux.

From the figure, it is evident that the contribution of the hot spot's emission in the short-wavelength region should vary significantly as the star rotates about its axis. Consequently, photometric observations in the $B$-band, for instance, should reveal a distinct periodicity for BP~Tau. However, no consensus has been reached on the period: different authors report values ranging from 7.6 to 8.3\,d (see, e.g., \citet{Wendeborn-II-2024, Lin-2023}). We also attempted to determine the star's rotation period by analysing radial velocity variations of emission lines in both the optical and UV spectra. The most precise measurements were obtained for the He\,I and He\,II lines. Nevertheless, the period could not be uniquely determined, as the observational data fit phase curves with similar scatter relative to a sinusoidal dependence across a range of periods between 7.0 and 8.5 days.

The absence of a clearly defined period suggests the presence of an additional factor modulating the radiation flux from the photosphere and hot spot system. This factor is likely the non-stationary nature of accretion, including irregular flares associated with the accretion process and/or coronal-chromospheric activity \citep{Costigan-2014}.  Incidentally, three flares lasting several tens of minutes were observed for BP\,Tau, during which the flux from the star varied by approximately 10\,\%. However, such flares are relatively rare events \citep{Lin-2023}.

\citet{Gotz-1961} reports that during the period rJD~35\,051--36\,200 (1954--1957), 
BP~Tau exhibited irregular variability, with brightness changes of $0\fm 6-0\fm 8$ often occurring over the course of a single day. Unfortunately, the context does not clarify whether the star's brightness increased or decreased during these events. Consequently, particular attention is drawn to the three deep dimming episodes of BP~Tau — episodes A, B, and C, described in Sect.~3.2. Recall that in episodes A and B, the dimming was accompanied by an increase in the $B-V$ colour index, which could, in principle, be attributed to a reduction in the flux contribution from the hot spot due to a decrease in the accretion rate.
  
However, the dimming event on December 26, 2013 (episode C), which was as deep as the previous ones but much shorter ($\Delta t \sim 1$\,h), occurred with a nearly constant $B-V\approx 0.8.$ In other words, as BP~Tau became fainter, it remained significantly blue, making it difficult to attribute the dimming to a reduced accretion rate. In our view, this dimming was likely caused by the obscuration of the hot spot by material in the accretion stream, while scattering of the star's light by dust particles in the infalling material made the light appear bluer, thereby compensating for the reduced contribution of the hot spot to the short-wavelength emission.

If the infall velocity of the material is $V_{\text ac} \sim 300$\,km\,s$^{-1}$ \citep{Wendeborn-III-2024}, then the extent of the gas-dust cloud eclipsing the star should be $l \approx V_{\text ac} \Delta t \sim  10^{11}$~cm, comparable to the stellar radius $R=1.5$\,R$_\odot$ \citep{Herzeg-Hillebrandt-2014}. This is consistent with the idea that the eclipse is caused by dimming due to a portion of the accretion stream projected onto the star. This suggests that the photometric variability of BP~Tau on time-scales ranging from several hours to several days may result not only from a variable accretion rate but also from variable circumstellar extinction.

\citet{Bouvier-1999} demonstrated that the periodic brightness variations of AA~Tau, accompanied by changes in the degree of polarisation \citep{Menard-2003}, are caused by the eclipse of the star by a gas-dust 'hump'. This hump forms at the inner edge of the protoplanetary disc due to its interaction with the stellar magnetosphere. Such phenomena have been observed in many CTTSs with discs viewed nearly edge-on \citep{Cody-2014}. Notably, this indicates that dust around late-type CTTSs can survive (without evaporating) in the accretion flow down to distances comparable to the stellar radius.
\footnote{Similar, but longer-lasting (on the order of several tens of days), eclipses may be caused by the so-called dusty disc wind -- see, e.g., \citet{Dodin-19}, \citet{Burlak-2024}.}

It is therefore reasonable to assume that the brightness variations of CTTS with discs significantly inclined to the line of sight may also result from variable extinction caused by dust falling onto the star. For such stars, including BP~Tau $(i\approx 38\degr),$ the brightness varies depending on the amount and size of dust reaching the region of the accretion stream projected onto the star. Quantitative modelling of BP~Tau's brightness variability, driven by variable extinction, is a non-trivial task. The solution depends on the properties of the dust particles in the region of interest (e.g., size distribution and scattering phase function), the geometry of the accretion stream, and the rotational phase.

It is equally important to understand the distribution of emission intensity within the hot spot, which remains unknown. The challenge lies in the fact that, in three-dimensional MHD simulations of accretion onto BP~Tau \citep{Long-2011}, the accretion rate is a free parameter of the problem. For the value chosen by the authors, $\dot M_{\text ac}<10^{-8}$\,M$_\odot$\,yr$^{-1}$, the maximum radiation flux from the surface of the hot spot does not exceed ${\cal F} \lesssim 10^{10}$\,erg\,s$^{-1}$cm$^{-2}$ (see, for example, Fig.~9 in their paper). This value is less than the photospheric flux corresponding to the effective temperature of BP~Tau, meaning that such an accretion spot cannot be significantly hotter than the photosphere. Moreover, modelling of BP~Tau's emission in the visible \citep{Dodin-13a, Wendeborn-III-2024} and UV \citep{Wendeborn-I-2024} ranges shows that, on average, $\log {\cal F} > 11$ within the spot.

\citet{Wendeborn-II-2024} analysed a large volume of photometric data obtained in 2021 and 2022 across various spectral bands in the visible range. They acknowledged that scattering by circumstellar dust might contribute to the star's brightness variability and even estimated the characteristic sizes of the dust particles required for such an effect. However, their analysis assumes that the dusty cloud uniformly obscures both the entire star and the hot spot simultaneously. Additionally, they did not consider how the nature of the eclipse might vary depending on the position of the hot spot relative to the observer or the angular dependence of its emission intensity (limb brightening). Notably, the constancy of the colour index during the dimming in episode C can be explained not only by the dominance of large ($\gtrsim 1 \mu$m) dust particles but also by high optical depth during a non-uniform eclipse caused by small dust grains \citep{1984ApJ...287..228N, 2021MNRAS.503.5704D}.

If variations in the dust content of BP~Tau's accretion stream are indeed a significant cause of its photometric variability, it is crucial to understand why the 'dustiness' of the infalling material changes. To address this, it is necessary to examine in detail how dust particles from the protoplanetary disc penetrate the star's magnetosphere and how long they can survive in the hot accretion stream. To observationally test this hypothesis, the first step is to gather data on how the polarisation of BP~Tau varies during the star's axial rotation.

We were only able to conduct polarisation observations of the star on three occasions. As shown in Table 2, the degree of polarisation in the $VRI$-bands was relatively small $(\sim 0.5\,\%{}),$ and the accuracy of our measurements does not allow us to determine whether the degree or angle of polarisation varies or if these parameters depend on wavelength. In the literature, we found only one reference to polarisation measurements of the star: \citet{Bastien-82} conducted observations on September 25 and November 14, 1976 using filters centred at 0.59 and 0.75\,$\mu$m. During these observations, the degree of polarisation was also $< 0.5\,\%{},$ and the observational errors were too large to draw any conclusions.

Let us now turn to the interpretation of the secular light curve of BP~Tau. In Sect.~3.1 we showed that the mean brightness of the star varies with an amplitude of $\Delta B \approx 0\fm 2$ and a characteristic time-scale of several decades. As seen in Fig.~\ref{fig:bp-lcurve}, after 1980, the increase in the star's mean brightness was accompanied by a decrease in the colour index $B-V$. This can be naturally explained by an increased contribution of radiation from the hot spot, corresponding to an enhanced accretion rate of material from the protoplanetary disc.

In principle, it cannot be ruled out that the accretion rate is modulated by the orbital motion of a low-mass companion to the star, such as a planet or brown dwarf. However, the absence of azimuthal asymmetry in BP~Tau's disc \citep{Jennings-2022}, combined with the constraints we derived on the companion's parameters (Sect.~3.3), makes this explanation less plausible. Alternatively, the secular variation in the accretion rate could be attributed to changes in the structure and/or strength of the star's magnetic field.

Radical changes in the character of photometric variability, apparently linked to variations in magnetic field parameters, have been observed in some T~Tauri-type stars, such as V410~Tau \citep{Yu-2019} and AA~Tau \citep{Bouvier-2013}. In the case of BP~Tau, the restructuring of the magnetic field structure may occur not abruptly, as in these stars, but smoothly, akin to the process during the Sun's 22-year activity cycle. According to \citet{Zaire-2024}, similar phenomena occur in GM~Aur: wavelike variations in the amplitude of brightness oscillations due to axial rotation $(P_{\text rot} \approx 6^{\text d})$ are associated with a non-stationary dynamo process, with a characteristic time-scale of $\approx 100$~days. It is worth noting that, according to \citet{Flores-2019}, the mean magnetic field strength of BP~Tau measured on November 6, 2017 was consistent within errors with the values reported by \citet{Donati-2008} for the beginning and end of 2006. However, this does not necessarily imply that the geometry of the magnetic field has remained unchanged over the 11 years.

In this context, we pay attention to the asymmetric structure of BP~Tau's bipolar jet. \citet{Dodin-2024} found that over the past approximately 150 years, Herbig-Haro objects have emerged exclusively in the receding part of the jet (the counter-jet), whereas previously, they appeared in the jet directed towards us. \citet{Kondratyev-2024} demonstrated that asymmetry between jets and counter-jets in supernova explosions can be explained by the lack of mirror symmetry of the magnetic field relative to the equatorial plane. As noted earlier, BP~Tau's magnetic field consists of a dipole and an octupole, which are inclined relative to each other and to the star’s rotation axis \citep{Donati-2008}, forming a configuration with broken mirror symmetry. If the counter-jet is currently the most active as a result, it is reasonable to assume that the magnetic field configuration was different approximately 150 years ago. In other words, the structure of BP~Tau's magnetic field has evolved over time.

Since the discussion has shifted to the jet of BP~Tau, we would like to point out the following circumstance. In the spectra obtained on January 5 and 7, 1988, \citet{Hartigan-1995} detected the [S\,II] 6716 and 6731~\AA{} lines, which, as is now understood, belong to the counter-jet. However, these lines were not visible in the spectrum obtained in October 1990 \citep{Hamann-1994}. As follows from  Fig.~\ref{fig:AB-dip}a, this discrepancy arises because the 1988 spectra were taken during a period when the star's brightness had significantly decreased, making the weak [S\,II] lines more prominent.


\section{Concluding remarks}
\label{sec:conclusion}

From the analysis of BP~Tau's light curve, we conclude that the parameters and locations of the hot (accretion) spots on the star's surface -- determined via magnetic Doppler imaging and corresponding three-dimensional MHD simulations -- are inconsistent with the lack of a clearly defined rotational modulation in the light curve. Most likely, this discrepancy arises from a combination of factors: the non-stationary nature of accretion, irregular eclipses caused by the dusty component of the accretion stream, and chromospheric and coronal activity.
ьной активности. 

In our view, all these factors are in some way linked to variations in the structure and/or strength of the star's magnetic field. It is possible that the variations in the mean brightness of BP Tau that we have detected, with characteristic time-scales of several decades, are also caused by long-term variations in the magnetic field.

Significant progress in understanding the causes of BP~Tau's non-trivial variability on different time-scales could be achieved through repeated magnetic Doppler imaging and polarimetric monitoring over two to three stellar rotations. These observations should be complemented by spectroscopic and photometric data gathered not only in the visible range but also in the near-infrared. In this context, this publication serves as a justification for the importance of such research.


\section*{Acknowledgments}
We dedicate this article to the memory of our recently deceased friends and colleagues Andrea Richichi and Christoph Leinert.

We would like to thank the staff of CMO SAI MSU for their assistance with the observations, Prof. W. Herbst for providing data from his photometric database \citep{Herbst-94}, and the referees for helpful comments. We also gratefully acknowledge the use of the SIMBAD database (CDS, Strasbourg, France), the Astrophysics Data System (NASA, USA), and the AAVSO database (https://www.aavso.org) in this work.

The work of A.V.~Dodin and B.S.~Safonov was supported by the RSF grant 23-12-00092. The results were obtained using equipment acquired as part of the development programme of Lomonosov Moscow State University.


\bibliographystyle{aspb1}
\bibliography{main}

\end{document}